\documentclass[letter]{ptptex}

\usepackage{amsmath}
\usepackage{graphicx}

\newcommand{\diff}[2]{\frac{\partial #1}{\partial #2}}

\newcommand{\lsim}{\raisebox{-0.55ex}{$\stackrel{\displaystyle <}{\sim}$}}

\title{
Description of $\eta$-distributions at RHIC energies in terms of the Ornstein-Uhlenbeck process and relativistic diffusion equations
}

\author{
Minoru \textsc{Biyajima}$^1$,  
Masahiro \textsc{Kaneyama}$^1$, \\
Takuya \textsc{Mizoguchi}$^2$
and Noriaki \textsc{Nakajima}$^3$ 
}

\inst{
$^1$Department of Physics, Shinshu University, Matsumoto 390-8621, Japan\\
$^2$Toba National College of Maritime Technology, Toba 517-8501, Japan\\
$^3$Center of Medical Information Science, Kochi University, Kochi, 783-8505 
Japan
}

\abst{
Assuming the Ornstein-Uhlenbeck process and relativistic diffusion equations 
with three sources at $y = 0$, we analyze $\eta$-distributions of $d$+Au 
collision as well as Au+Au collisions at $\sqrt{s_{NN}} =$ 130 GeV and 200 GeV.
}

\begin{document}

\maketitle

\begin{figure}
    \centerline{\includegraphics[height=7.0 cm]{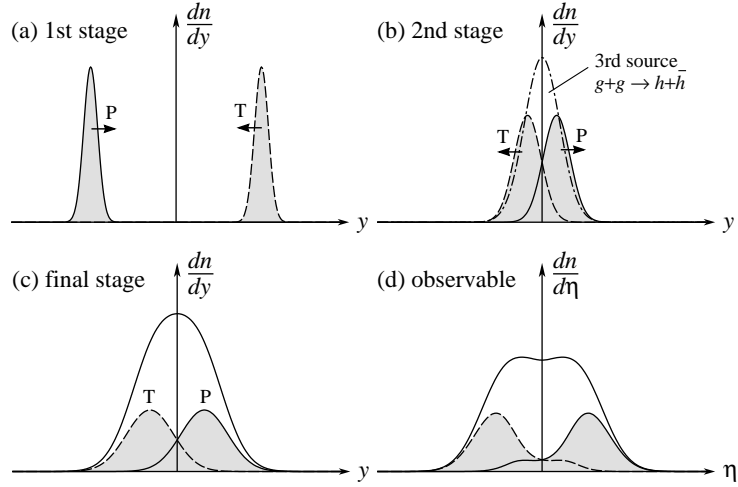}}
\caption{Description of two-step processes by relativistic diffusion equations 
and the Ornstein-Uhlenbeck process.}
\label{fig1}
\end{figure}

\begin{figure}
    \centerline{\includegraphics[height=7.0 cm]{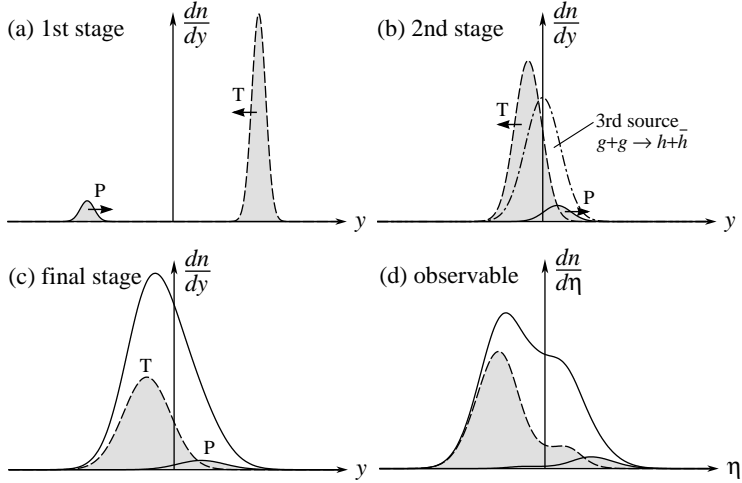}}
\caption{Description of two-step processes for $d$+Au collisions.}
\label{fig2}
\end{figure}

\noindent
\textbf{1 Introduction}: 
$\eta$-distributions in Au+Au collisions at RHIC with various centrality-cuts 
(C.C.) at RHIC energies have been reported in Refs.~\citen{Back:2001bq} and 
\citen{Nouicer:2002ks}. In Ref.~\citen{Biyajima:2003nv} we have proposed a stochastic model with the sources at $\pm \, y_{max}$ and $y = 0$, to describe 
the $\eta$-distributions. That is an extended model from a model of two sources 
at $\pm y_{max}$ in Refs.~\citen{Biyajima:2002at} and 
\citen{Biyajima:2002wq}~\footnote{
In Ref.~\citen{Biyajima:2002wq}, we has pointed out that single Gaussian distribution with an initial condition $\delta (y)$ cannot explain the whole 
$\eta$-distributions in Au+Au collisions. This fact suggests that 
contributions from the projectile and target fragmentations are necessary.
}. 
In this letter, we propose a model with three sources at $y=0$. A more realistically physical picture is shown in  Figs.~\ref{fig1} and \ref{fig2}, 
where our basic assumption is shown therein.
  
Very recently PHOBOS Collaboration has reported $\eta$-distribution on $dN/d\eta$
 ($d$ + Au) at $\sqrt{s_{NN}} = 200$ GeV~\cite{Back:2003hx}. Thus we are interested in analyses of the data, to examine whether or not the model proposed 
 in Figs.~\ref{fig1} and \ref{fig2} works well for descriptions of $dN / d \eta$ 
 ($d$+Au) as well as $dN / d \eta$ (Au+Au). 
%
%
%
%

\noindent
\textbf{2 Theoretical formula}: 
To improve the physical picture of Ref.~\citen{Biyajima:2003nv}, we consider a 
formula based on the stochastic approaches. To describe physical picture of 
Figs.~\ref{fig1} and \ref{fig2}(b), we have found that the following 
relativistic diffusion equations are available from studies of 
Ref.~\citen{Back:2003hx},
\begin{eqnarray}
  \diff{P(y,\, t)}t = \gamma \left[\diff{}y(y\pm y_{max}) + 
  \frac 12\frac{\sigma^2}{\gamma}\diff{^2}{y^2}\right] P(y,\, t)\:.
\label{eq:1}
\end{eqnarray}
Using the initial condition
\begin{eqnarray}
  P(y,\, 0) = \delta (y)
\label{eq:2}\:,
\end{eqnarray}
we obtain the following solutions for the projectile and target fragmentations
\begin{subequations}
\label{eq:3}
\begin{eqnarray}
  P^{(T)}(y,\, y_{max},\, t) &=& \frac 1{\sqrt{2\pi V_T^2(t)}}
  \exp\left[-\frac{(y-y_{max}(1-e^{-\gamma t}))^2}{2V_T^2(t)}\right]
 \label{eq:3a}\:,\\
  P^{(P)}(y,\, -y_{max},\, t) &=& \frac 1{\sqrt{2\pi V_P^2(t)}}
  \exp\left[-\frac{(y+y_{max}(1-e^{-\gamma t}))^2}{2V_P^2(t)}\right]
 \label{eq:3b}\:,
\end{eqnarray}
\end{subequations}
where $V_{(P\ {\rm or}\ T)}^2 = (\sigma^2/2\gamma)(1-e^{-2\gamma t})$. For the 
3rd source due to gluons in the central region, we assume the Ornstein-Uhlenbeck 
process (Eq.~(\ref{eq:1}) without the term $y_{max}$; see 
Ref.~\citen{Biyajima:2003nv}). The following solution can be derived from 
Eq.~(\ref{eq:2}) as 
\begin{eqnarray}
  P^{(C)}(y,\, t) = \frac 1{\sqrt{2\pi V_0^2(t)}}
  \exp\left[-\frac{y^2}{2V_0^2(t)}\right]
\label{eq:4}\:.
\end{eqnarray}
Combining Eqs.~(\ref{eq:3}) and (\ref{eq:4}), we obtain the following expression
\begin{eqnarray}
\frac{dn}{d\eta} = J(\eta, \, m/p_{t}) \times &&\left\{\frac{c_T}{c_T+c_P+c_0} 
P^{(T)}(y,\, y_{max}, \, t)\right.
\nonumber\\
&&\left. + \frac{c_P}{c_T+c_P+c_0} P^{(P)}(y,\, -y_{max}, \, t)\right\} 
\nonumber\\
&&+ J_0\left\{\frac{c_0}{c_T+c_P+c_0} P^{(C)}(y, \, t)\right\}
\label{eq:5}\,,
\end{eqnarray}
where $c_T$, $c_P$ and $c_0$ are the weight factors of three contributions for 
the target fragmentation, the projectile one and the central region, 
respectively. A relation between $y$ and $\eta$, and the Jacobians are given as 
\begin{equation}
y = \frac 12\ln \frac{\sqrt{1+(m/p_t)^2 + \sinh^2 \eta} + 
\sinh \eta}{\sqrt{1+(m/p_t)^2 + \sinh^2 \eta} - \sinh \eta}
\label{eq:6}\:,
\end{equation}
\begin{equation}
J(\eta, \, m/p_t) = \frac{\cosh \eta}{\sqrt{1+(m / p_t)^2 + \sinh^2\eta}}
\label{eq:7}\:,
\end{equation}
and $J_0 = J(\eta, \: m_0/p_t)$. 

\begin{figure}
    \centerline{\includegraphics[height=6.0 cm]{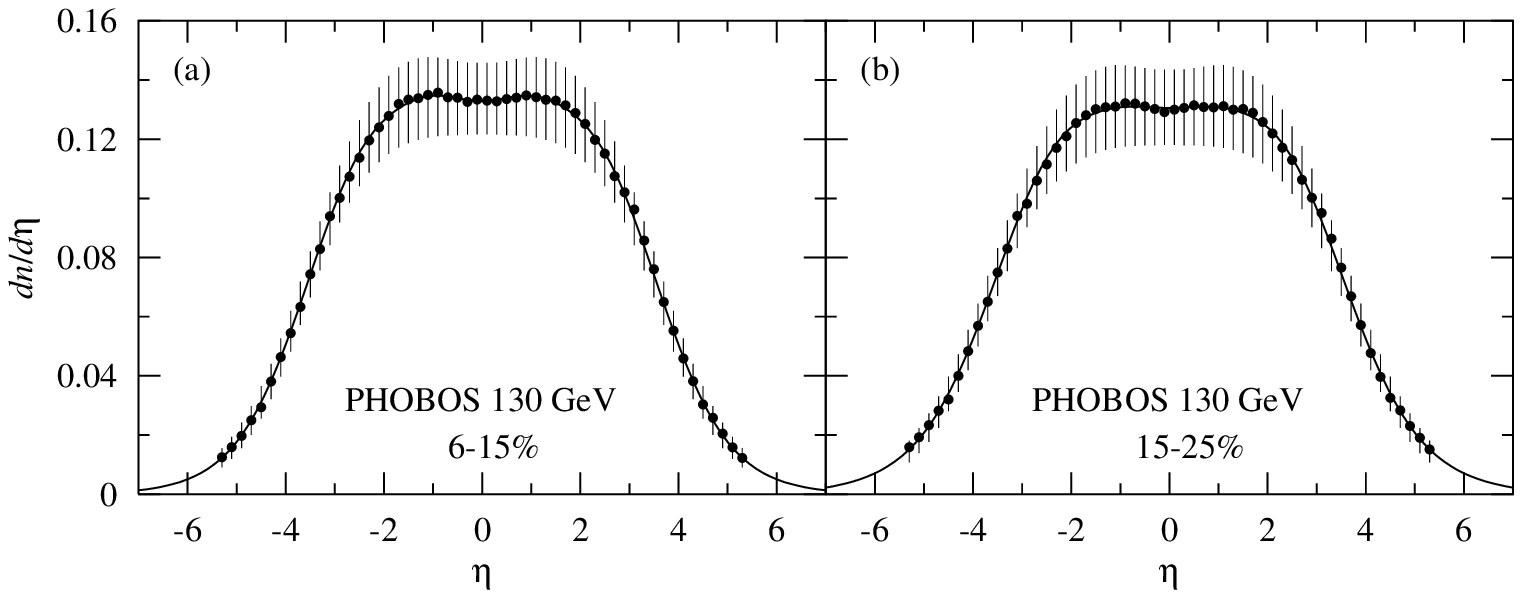}}
    \centerline{\includegraphics[height=6.0 cm]{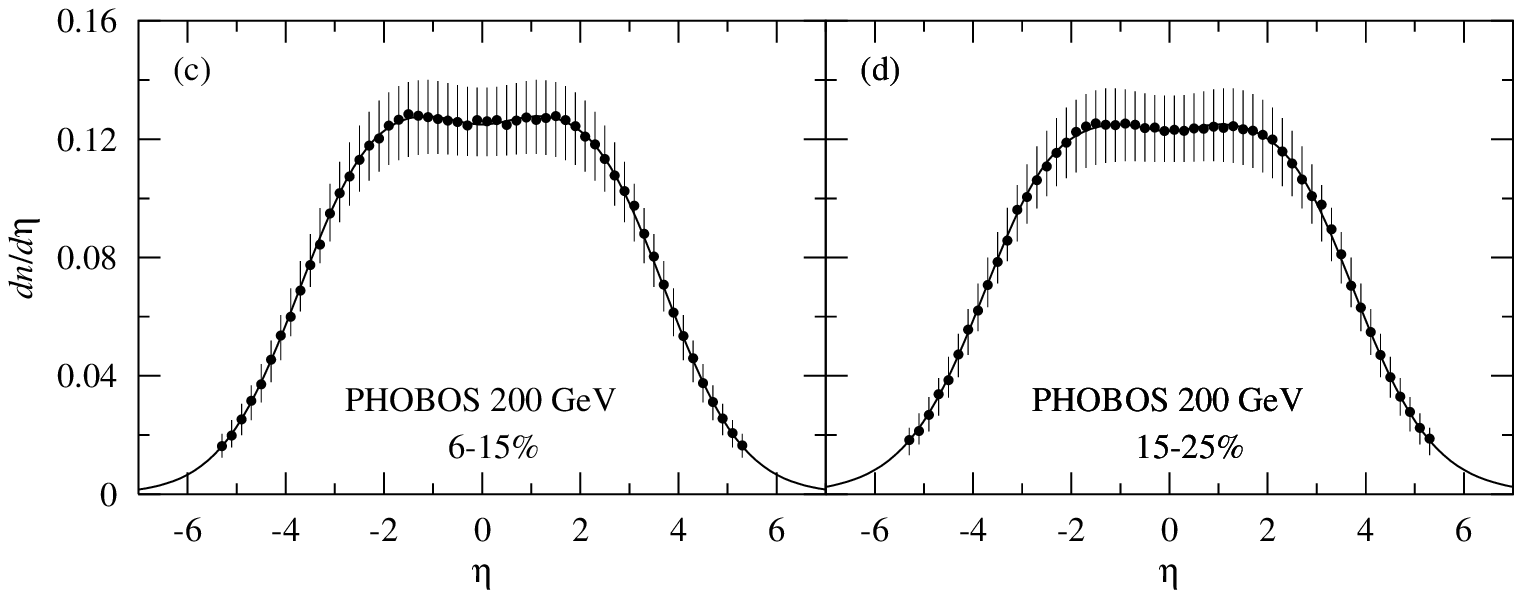}}
\caption{Analyses of data on $dn/d\eta = (N_{ch})^{-1}dN_{ch}/d\eta$ at 
$\sqrt{s_{NN}} =$ 130 GeV and 200 GeV. $C.C.=$6-15 \% and 15-25 \%.}
\label{fig3}
\end{figure}

\noindent
\textbf{3 $\eta$-distributions in Au+Au collisions at smaller centrality-cuts 
(C.C.)}: 
We apply Eq.~(\ref{eq:5}) to the data on $\eta$-distributions at the C.C.6-15\% 
and 15-25\% at $\sqrt{s_{NN}} =$ 130 GeV and 200 GeV by the PHOBOS 
Collaboration. Our results, using the CERN MINUIT program, are shown in 
Fig.~\ref{fig3} and Table~\ref{table1}. We have to determine the weight factors, 
$c_P$, $c_T$ and $c_0$. For this aim, we adopt the following criterions 
$m_0/p_t > \delta (m_0/p_t)$, $m/p_t > \delta (m/p_t)$ and $m_0/p_t < \delta 
m/p_t$, because pions are mainly produced in the central region. They reveal 
that the ratio is $c_P : c_0 : c_T = 1 : 6$-$7 : 1$ at the minimum $\chi^2$'s, 
where the evolution time $p= (1-e^{-2\gamma t})$ stops.
%
%
%
%
\begin{table}
\begin{center}
\caption{Values of estimated parameters of Eq.~(\ref{eq:5}) with $c_P/c_T = 1$ 
for Au + Au at $\sqrt{s_{NN}} = 130$ GeV and 200 GeV.}
\label{table1}
\begin{tabular}{ccccc} \hline\hline
               & \multicolumn{2}{c}{130 GeV}   & \multicolumn{2}{c}{200 GeV}\\
C. C.          & 6-15\%        & 15-25\%       & 6-15\%        & 15-25\%\\
\hline
$p$            & 0.52$\pm$0.21 & 0.62$\pm$0.13 & 0.78$\pm$0.12 & 0.66$\pm$0.16\\
$V_0^2$        & 5.17$\pm$0.95 & 5.98$\pm$1.03 & 5.64$\pm$1.42 & 6.06$\pm$1.67\\
$V_T^2$        & 1.31$\pm$0.26 & 1.34$\pm$0.30 & 1.43$\pm$0.25 & 1.59$\pm$0.28\\
$c_0$          & 2465$\pm$39   & 1808$\pm$32   & 3078$\pm$54   & 2120$\pm$40\\
$c_P$          & $c_0/6$       & $c_0/7$       & $c_0/7$       & $c_0/6$\\
$m_0/p_t$      & 0.44$\pm$0.26 & 0.37$\pm$0.29 & 0.47$\pm$0.33 & 0.42$\pm$0.33\\
$m/p_t$        & 3.60$\pm$2.30 & 2.34$\pm$1.06 & 1.06$\pm$0.88 & 2.06$\pm$1.14\\
$\chi^2$/d.o.f.& 0.64/48       & 0.82/48       & 0.50/48       & 0.54/48\\
\hline
\end{tabular}
\end{center}
\end{table}

\begin{figure}
    \centerline{\includegraphics[height=6.0 cm]{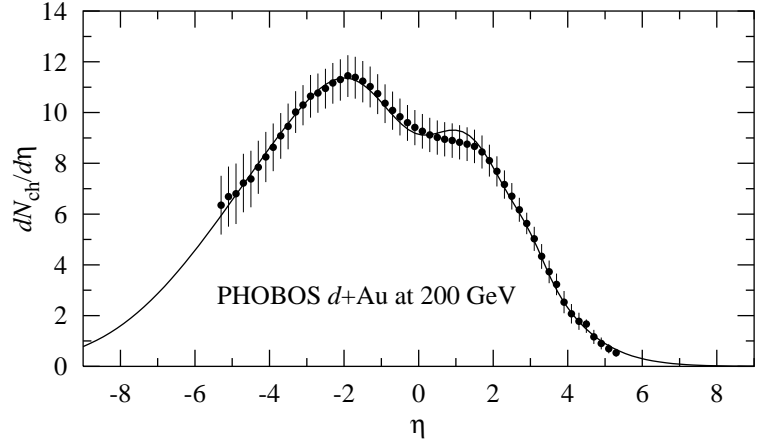}}
\caption{Analyses of data on $d$+Au collision at $\sqrt{s_{NN}} =$ 200 GeV by 
Eq.~(\ref{eq:5}).}
\label{fig4}
\end{figure}

\noindent
\textbf{4 $\eta$-distributions in $d$+Au collision}: 
Taking into account the mass ratio, $m_d : m_{Au} = 1 : 100$, we can fix the 
ratio as $c_T/c_P = 100$. The weight factor $c_0$ is treated as a free parameter 
in the present analysis. In eq.~(\ref{eq:5}), two Jacobian factor, $J(\eta, \, m_T/p_t)$ and $J(\eta, \, m_P/p_t)$ are used. The evolution parameter $p=1-e^{-2\gamma t}$ is also 
stopped at minimum $\chi^2$. Our result is given in Fig.~\ref{fig4} and 
Table~\ref{table2}. We know that the weight factors are described by 
$c_P : c_0 : c_T = 0.53 : 40 : 53$. This ratio means contribution from the gluon 
is the same magnitude of the target fragmentation. 
%
%
%
%
\begin{table}
\begin{center}
\caption{Parameter values of Eq.~(\ref{eq:5}) with $c_T/c_P = 100$ for $d$+Au at 
$\sqrt{s_{NN}} = 200$ GeV~\cite{Back:2003hx}.}
\label{table2}
\begin{tabular}{ccccc} \hline\hline
$p$ & $V_0^2$ & $V_P^2$ & $V_T^2$ & $c_0$\\
 0.76 &
  4.27$\pm$0.64 &
 0.165$\pm$0.159 &
  7.56$\pm$1.38 &
  39.8$\pm$5.4 \\
\hline
$c_P$ & $m_0/p_t$ & $m_P/p_t$ & $m_T/p_t$ & $\chi^2$/d.o.f.\\
 0.529$\pm$0.077 &
 0.556$\pm$0.166 &
  1.63$\pm$0.69 &
  1.88$\pm$0.41 &
  5.1/45\\
\hline
\end{tabular}
\end{center}
\end{table}

\noindent
\textbf{5 Concluding remarks}: 
\begin{enumerate}
  \item To improve the previous physical picture in Ref.~\citen{Biyajima:2003nv},
   we assume the relativistic diffusion equations for the target and projectile 
   fragmentations in addition to the Ornstein-Uhlenbeck stochastic process available for that in the central region.
  \item The ratio, $c_P : c_0 : c_T$, is determined by means of the following 
  criterions; $m_0/p_t > \delta (m_0/p_t)$, $m/p_t > \delta (m/p_t)$ and 
  $m_0/p_t < \delta m/p_t$. They can be investigated by analyses of 
  $\eta$-distributions with restrictions $\eta_{c_1} \lsim |\eta| \lsim 
  \eta_{c_2}$ in a future.
  \item As seen in Figs.~\ref{fig3} and \ref{fig4}, our formula seems to be 
  available for the description of $\eta$-distributions in $d$+Au as well as 
  Au+Au collisions.
  \item From analysis of $d+$Au collision, $c_P : c_0 : c_T = 0.5 : 40 : 53$, 
  we have know that the contribution of gluon in Au is the same ordered 
  magnitude as that of quarks.
  \item Since our formula is expressed in the analytical form, it can be applied 
  to other reactions, for example, Si+Al and S+Au collisions. These analyses 
  will be shown elsewhere~\cite{Biyajima:2004aa}.\\
\end{enumerate}

\noindent
\textbf{Acknowledgements}:
We would like to thank R.~Nouicer for his kind information on $dn/d\eta$ ($d$+Au), and discussions with N.~Suzuki. One of authors (K.M.) is partially supported by Shinshu University.

\end{document}